\documentclass[11pt]{article}
\usepackage{a4}
\usepackage{amssymb}
\usepackage{graphicx}
\usepackage{hyperref}
\newcommand{\be}{\begin{equation}}
\newcommand{\ee}{\end{equation}}
\newcommand{\bea}{\begin{eqnarray}}
\newcommand{\eea}{\end{eqnarray}}
\newcommand{\ps}{p_m\sigma^m}
\newcommand{\qs}{q_m\sigma^m}

\begin{document}

\title{Higher loop renormalization of a supersymmetric field theory}
\author{Marc P. Bellon$^{a,b}$\thanks{On leave from,
Laboratoire de Physique Th\'eorique et Hautes Energies, Boite 126,
4 Place Jussieu,  75252 Paris Cedex 05. Unit\'e Mixte de Recherche
UMR 7589 Universit\'e Pierre et Marie Curie-Paris6; CNRS;
Universit\'e Denis Diderot-Paris7.}~, Gustavo Lozano$^c$ and Fidel
A.\ Schaposnik$^{b,a}$\thanks{Associate to CICBA.}\\
 {\normalsize \it $^a$CEFIMAS, Av.\,Santa Fe 1145,
 1069 Capital Federal,
Argentina}\\
{\normalsize\it $^b$Departamento de F\'\i sica,
Universidad Nacional de La Plata}\\ {\normalsize\it C.C. 67, 1900
La Plata, Argentina}
\\
 {\normalsize \it$^c$Departamento de F\'\i sica,
FCEyN, Universidad de Buenos Aires}\\ {\normalsize\it Pab.I,
Ciudad Universitaria, 1428, Buenos Aires, Argentina}}
\date{\hfill}
\maketitle
\begin{abstract}Using Dyson--Schwinger equations
within an approach developed by Broadhurst and Kreimer and the renormalization
group,
 we show how high loop order of the
renormalization group coefficients can be efficiently computed in a
supersymmetric model.
\end{abstract}

\section{Introduction}
Perturbative Quantum Field Theory is known for its tremendous
successes, with its ability to obtain highly precise values in
Quantum Electrodynamics, or the tests of the standard model involving
radiative corrections to weak interactions. However, actual calculations
become rapidly cumbersome and display a conjunction of analytical and
combinatorial difficulties combined with the very fast growth of the
number of relevant terms, which get compounded by the need to address
the subtraction of the various subdivergencies.

In the quest of organizing principles for taming the combinatorial
problem, a major progress has been the recognition of a Hopf algebra
structure in the renormalization of Quantum Field Theories
\cite{kreimer1} and subsequent developments \cite{CK1}-\cite{CK3}
and the link relating the cohomology of the introduced Hopf algebras
with Dyson-Schwinger equations \cite{BrKr02}-\cite{BrKr01} (See
\cite{Krev} for a complete list of references). An application to
the summation of a category of diagrams in a simple Yukawa field
theory or a $\phi^3$ theory in 6 dimension has been obtained by
Broadhurst and Kreimer in~\cite{BrKr01}. In that work a very
powerful method to compute higher order corrections to the anomalous
dimension has been developed. Basically, it is based on a
differential equation derived from the Dyson-Schwinger equation for
the self-energy.

The purpose of this letter is to extend such results to the case of
a supersymmetric model. We have considered a Wess--Zumino like model
in which the non-renormalization of the vertex ensures that the
order zero in a large $N$ expansion of the renormalization group
functions can be obtained.  We also expect that the known
non-renormalization properties related to supersymmetry should allow
to go beyond this initial category of diagrams in future works.
Along the way, we propose an alternative derivation of the
propagator-coupling duality presented in~\cite{BrKr01} based on
intuitive renormalization group arguments.

\section {The model}

We consider massless chiral superfields $\Psi$ and $\Phi_i$
($i=1,2,\ldots, N$) and their (antichiral) complex conjugates
$\Psi^+$ and $\Phi_i^+$,
which satisfy the following constraints:%
\begin{eqnarray}
\bar D_{\dot \alpha} \Psi= 0 \, , \;\;&&\;\; \bar D_{\dot \alpha} \Phi_i= 0 \nonumber\\
 D_{ \alpha}   \Psi^+= 0 \, , \;\;&&\;\;   D_{ \alpha} \Phi_i^+= 0
\end{eqnarray}
The derivatives are defined as:
\begin{equation}
D_\alpha = \frac{\partial}{\partial \theta^\alpha} +
 2i\sigma_{\alpha \dot\alpha}^\mu{\bar \theta}^{\dot\alpha}
 \frac{\partial}{\partial y^\mu} \; , \;\;\; \;\;\; \;\;\;
 \bar D_{\dot \alpha} = -\frac{\partial}{\partial {\bar\theta}^{\dot\alpha}}
\end{equation}
in terms of the chiral coordinates $y$ defined as
\begin{equation}
y^\mu = x^\mu + i \theta \sigma^\mu\bar \theta
\end{equation}

Each chiral superfield represents a complex scalar ($A, B_i$),
a Weyl fermion ($\chi, \xi_i$) and a complex auxiliary field ($F,G_i$)
as it can be seen from their expansion in the $\theta$ variables,
\begin{eqnarray}
\Psi &=& A(y) + \sqrt 2 \theta \chi(y) + \theta \theta F(y) \nonumber
\\
\Phi_i &=& B_i(y) + \sqrt 2 \theta \xi_i (y) + \theta \theta G_i(y)
\label{uno}
\end{eqnarray}

Dynamics is governed by the Lagrangian density
\begin{equation}
L = \int d^4\theta \Psi \Psi^+ +
\sum_{i=1}^N \int d^4\theta \Phi_i \Phi_i^+
+
\frac{g}{\sqrt N} \sum_{i=1}^N \int d^2\theta \Phi_i\Psi^2
\end{equation}
The cubic superfield interaction corresponds to a Yukawa coupling
between scalar and fermions and, after the elimination of the
auxiliary fields, to a quartic coupling of the scalars. The
Lagrangian reads in terms of component fields:
\begin{eqnarray}
L &=& i\partial_\mu \bar \chi \bar \sigma^\mu \chi +
i\sum_{i=1}^N \partial_\mu \bar \xi_i \bar \sigma^\mu \xi_i
+ A^* \Box A + \sum_{i=1}^N  B_i^* \Box B_i +
F^*F + \sum_{i=1}^N G_i^*G_i
\nonumber\\
&& + \frac{g}{\sqrt N} \sum_{i=1}^N \left(
 A^2 G_i +2 A B_i F- \chi^2 B_i - 2 \xi_i\chi A +{\rm h.c.}
\right)
\end{eqnarray}
In our conventions (the same as in \cite{WB}), we have ${\rm diag}\;
g_{mn} = (-1,1,1,1)$ and
\begin{equation}
\Box = -\partial_t^2 + \nabla^2
\end{equation}

\section{Dyson--Schwinger equations}

We only consider corrections to the propagators of the fields
associated to the superfield $\Psi$, as corrections to the
propagation of the superfield $\Phi_i$ get a factor $1/N$ and do not
contribute in the large $N$ limit. Moreover supersymmetry ensures
that there are no divergent vertex corrections so that what we are
calculating represents effectively the order zero of the large $N$
development.

The set of diagrams that we consider is the same as the one studied
in~\cite{BrKr01}, except that now the diagrams have to be considered
as superfield diagrams and therefore, each one will correspond to a
sum of ordinary Feynman diagrams. The number of such diagrams grows
rapidly with the number of loops and for example a given 12 loop
superdiagram corresponds to 18144 ordinary diagrams.

The diagrams we are looking at correspond to the expansion of the
Dyson--Schwinger equations graphically represented in Figure 1.

With our (Minkowski) conventions, free propagators read
\begin{eqnarray}
\Pi_{0A}^{-1}(p) &=& \langle 0 \vert T\left( A(x)
A^*(x')\right)\vert 0\rangle = i\,
\Box^{-1}(x-x')\nonumber\\
\Pi_{0F}^{-1}(p)&=& \langle 0 \vert T\left( F(x) F^*(x')\right)\vert
0\rangle =
  i\,
\delta(x-x')\nonumber\\
\left(\Pi_{0\chi }^{-1}\right)_{\!\!\alpha \dot \beta}\!\!(p) &=& \langle 0 \vert T\left( \chi_\alpha(x)
{\bar\chi}_{\dot\beta}(x')\right)\vert 0\rangle = \sigma^m_{\alpha
\dot\beta}  \partial_m \Box^{-1}(x-x')
\end{eqnarray}

\begin{figure}
\centering
\includegraphics[width=15cm]{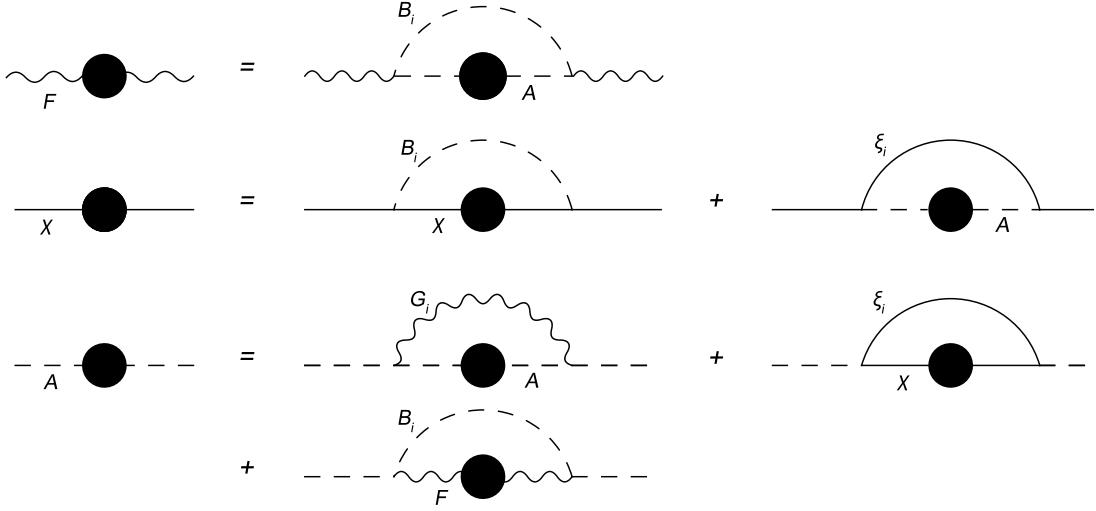}
\caption{\small Schwinger-Dyson equations} \label{F12T2Lat}
\end{figure}

It will be convenient to write Schwinger-Dyson equations in
Euclidean space. Then calling $\Pi_F$, $\Pi_\chi$, $\Pi_A$ the full
propagators for fields $F, \chi$ and $A$ in (Euclidean) momentum
space, the corresponding self-energies are introduced as
\begin{eqnarray}
\Pi_A^{-1}(p) &=& p^2 \left(1 - \Sigma_A(p^2)\right)\nonumber\\
\Pi_F^{-1}(p) &=& -(1 - \Sigma_F(p^2)) \nonumber\\
\Pi_\chi^{-1}(p) &=& p_m\sigma^m \left(1 - \Sigma_\chi(p^2)\right)
\end{eqnarray}
We consider a simple one-loop correction to the self-energy,
where one of the propagators has a momentum-dependent renormalization
$\Sigma(p^2)$.
The simplest case is the one for the self-energy of the auxiliary field $F$,
since there are only contributions from a single diagram with scalar
propagators:
\begin{equation} \label{sigma}
    \Sigma_F(p^2) = -\frac{g^2}{4\pi^4}\int d^4q {1\over q^2 (1-\Sigma_A(q^2)) (p-q)^2}
\end{equation}
For the fermion $\chi$, we have the two contributions shown in figure~1,
\begin{eqnarray}
\ps \, \Sigma_\chi(p^2) &=&  -\frac{g^2}{4\pi^4}\left(
\int d^4q \frac{\qs}{q^2(1- \Sigma_\chi(q^2))(p-q)^2} + \right.\nonumber\\
&& \left.\int d^4q \frac{\ps - \qs}{q^2(1- \Sigma_A(q^2))(p-q)^2}
\right)\label{sigmaChi}
\end{eqnarray}
In the case of the scalar $A$, the three different terms shown in figure~1
evaluate as:
\begin{eqnarray}
p^2 \Sigma_A(p^2) &=&  \frac{g^2}{4\pi^4} \left(
\int d^4q \frac{-1}{q^2(1- \Sigma_A(q^2))}
+ \int d^4q \frac{-1}{(1- \Sigma_F(q^2))(p-q)^2}\right.
\nonumber\\
&+&\left.
\int d^4q \frac{-{\rm Tr}\bigl(q_n\bar\sigma^n(\qs-\ps) \bigr)}
{q^2(1- \Sigma_\chi(q^2))(p-q)^2} \right)\label{sigmaA}
\end{eqnarray}

A consistent ansatz to solve these equations is to take
\begin{equation}
\Sigma_A = \Sigma_\chi = \Sigma_F = \Sigma
\end{equation}
Indeed, with this ansatz, the two integrals in~(\ref{sigmaChi}) combine
to give $\ps$ times the right hand side of~(\ref{sigma}). Similarly,
the equation for $\Sigma_A$~(\ref{sigmaA}) becomes $p^2$
times~(\ref{sigma}). Note that ${\rm Tr}(\sigma_n \bar\sigma_m) =
-2 g_{mn}$.

The integration in (\ref{sigma}) can be done along the same lines
as in~\cite{BrKr01}. The angular integration uses the fact that
the angular average of $1/(p-q)^2$ is $1/\max(p^2,q^2)$. With the
variables $x=p^2$ and $y=q^2$, the radiative correction becomes
\begin{equation}
    \Sigma(x) = -\frac{g^2}{4\pi^2} \Bigl[ \int_0^x dy {1\over x(1-\Sigma(y))}
 + \int_x^\infty {1\over y(1-\Sigma(y))} \Bigr]
\end{equation}
Taking the $x$ derivative to get rid of the infinite constant, we
arrive first at
\begin{equation}
    x^2 {d\over dx}\Sigma(x) = +\frac{g^2}{4\pi^2} \int_0^x dy {1\over (1-\Sigma(y))}
\end{equation}
A further derivation allows to obtain a simple differential equation
for $\Sigma$,
\begin{equation} \label{resume}
    (1-\Sigma(x))D(D+1) \Sigma(x) =  \frac{g^2}{4\pi^2} = a
\end{equation}
with $D= x d/dx$, which is equivalent to $D=d/d \log(p^2/\mu^2)$.

\section{Anomalous dimension}
The equation~(\ref{resume}) is very similar to the one obtained
for the Yukawa case in~\cite{BrKr01}. The only difference is that
the operator $D(D+1)$  replaces in the present case the  $D(D+2)$
one in~\cite{BrKr01}. The two equations can then be identified by
rescaling the variable $L=\log(p^2/\mu^2)$ and $a$ by appropriate
factors. The results derived   in~\cite{BrKr01} can then be
reobtained taking into account  the correct  factors.

The most direct way of obtaining a solution of eq.~(\ref{resume}) is
to solve for $\Sigma(x)$ order by order in $a$,   using the
renormalization constraint $\Sigma(\mu^2)=0$.
 At order $n$ in $a$, $\Sigma(x)$ is a polynomial in $L$ of
order $n$ without constant terms. The renormalization group $\gamma$
function is then the first derivative in $L$ for $L=0$.
\begin{equation}\label{}
 \gamma(a)= \left. \frac{d\log(1 -\Sigma(p^2,a))}{d\log(p^2)}
\right\vert_{p^2=\mu^2}
\end{equation}
Here $\gamma$ is the anomalous dimension for all the fields in the
scalar multiplet $\Psi$.

As was shown already in~\cite{BrKr01},  this procedure can be
further simplified if one directly computes $\gamma$ through the use
of a non-linear equation that can be deduced from
eq.~(\ref{resume}). Indeed, consider the renormalization group
relation~\cite{bogo}
\begin{equation}
d\left( \frac{p^2}{\mu^2}, a\right) = d\left( \frac{p^2}{s^2},a(s^2)\right)
d\left( \frac{s^2}{\mu^2},a\right)
\label{b}
\end{equation}
where
\begin{equation}
d(p^2,a)^{-1} = 1 - \Sigma(p^2,a)
\end{equation}
Differentiation of eq.~(\ref{b}) with respect to $p^2$ one obtains
\begin{equation}
\frac{d\log(1-\Sigma)}{d\log p^2} =  \gamma(a(s^2))
\label{rela1}
\end{equation}
or
\begin{equation}
\frac{d\log(1-\Sigma)}{d\log p^2} =
 \gamma\left(\frac{a}{(1-\Sigma)^2}\right)
 \label{rela}
\end{equation}
This last identity  was obtained as a consequence of the
non-renormalization of the vertex which implies a connection between
the $\beta$ and $\gamma$ functions of the renormalization group. It
should be noted that eq.~(\ref{rela}), which was derived here using
the renormalization group, can also be derived using Hopf algebra
arguments as explained in~\cite{BrKr01}. Indeed, in this last work
the dependence of the self--energy on the momentum and the anomalous
dimension (the propagator--coupling duality) based on the Hopf
structure of Feynman diagrams is used to prove eq.~(\ref{rela}).
Here, our derivation relies instead on the renormalization group
relation eq.~(\ref{b}).

Let us now rewrite eq.~(\ref{rela})  in the form
\begin{equation}
 D(1-\Sigma) = (1-\Sigma)\gamma(a/(1-\Sigma)^2)
\end{equation}
 and apply the operator $(D+1)$ to both sides. After some work  one ends with
\begin{equation}
 \gamma = -a + a^2 \frac{d(\gamma^2/a)}{da}
 \label{nonlinear}
\end{equation}
From this nonlinear differential equation one can derive the perturbative
expansion of $\gamma$,
\begin{eqnarray}
\gamma &=& \sum_{n>0} G_n  (-a)^n \nonumber\\
G_{n+1} &=& n \sum_{k=1}^{n} G_k G_{n+1-k}
\label{recu}
\end{eqnarray}

A simple Mathematica program allows to evaluate $G_n$ for large values of $n$
according to this recurrence relation. As an example, $G_{100}$ can be evaluated
quasi immediately giving
\begin{eqnarray}
G_n &=&
2421240422948166903917318482712165846373737073005111621110\nonumber\\
&&
3204862084235299310224396534696102431673533019446585512533\nonumber\\
&&
0800645887484858421550031445472180813110312557469722336097\nonumber\\
&&3401051786164
\end{eqnarray}

We can then compute $\gamma$ as a function of $a$ using eq.~(\ref{recu}).
However
being  $G_n \sim (2n-1)!!$ for large $n$, the series is  asymptotic
and therefore we use a Pad\'e--Borel resummation method. We start
from the approximation
\begin{equation}
\gamma \approx  - a \int dx  P(ax) \sqrt x \exp(-x)
\end{equation}
where $P(x)$ is defined by the following series
\begin{equation}
P(x) = \sum(-1)^n \frac{G_nx^{n-1}}{\Gamma(n + 1/2)}
\end{equation}
Numerical values for $\gamma$ can then be obtained using a Pad\'e
$[N\backslash M]$ approximant for $P(x)$. If one avoids the values
of $[N\backslash M]$ for which this approximant has poles on the
positive real axis, one can obtain stable evaluations for growing
values of $a$. The successive approximations are fully coherent and
higher degree approximants allow to reach higher values of the
coupling constant $g$, $a = g^2/4\pi^2$. In a few seconds one can
compute the anomalous dimension $\gamma$ for values  $g^2 \approx
700$ with a relative accuracy of the order of $10^{-10}$. As an
example, one finds $\gamma(g^2 = 789.568) =-6.30706 $ with a
$[99\backslash100]$ approximant. Rational solutions are not possible
for such high orders and the precision afforded by floating point
numbers is not sufficient to obtain solutions, but the use of higher
precision allowed to reach easily this stage.

In figure~2, we show $\gamma$ as a function of $g^2$. Remarkably, we
see from the numerical results that the anomalous dimension $\gamma$
of the scalar field can reach values where it exceeds the canonical
one, so that, unexpectedly  the propagator grows with the impulsion.
One confirms this analytically from  the asymptotic behavior of
$\gamma$ for large $g$ that can be easily inferred from
eq.~(\ref{nonlinear}),
\begin{equation}\label{asymp}
    \gamma \approx g \sqrt{\ln g}
\end{equation}

We see that  our calculation gives a complete knowledge of the
renormalization group functions since, due to the
non-renormalization of the vertex one has $\beta  = -2\gamma$. In
particular, the asymptotic behavior of $\gamma$ in eq.~(\ref{asymp})
allows to prove that the coupling constant goes to infinity for a
finite value of the scale.

We end this section by noting that the renormalization group played
a central r\^ole in obtaining the basic equations
(\ref{rela})-(\ref{nonlinear})  from the Dyson--Schwinger equations.
General information on the renormalization group have been presented
in~\cite{CK3}, deduced from the Hopf algebra structure of perturbative
quantum field theory. It would be of interest to combine the two approaches
to fully realize the potential of Dyson--Schwinger methods to obtain renormalization
group functions.

\begin{figure}
\centering
\includegraphics[width=13cm]{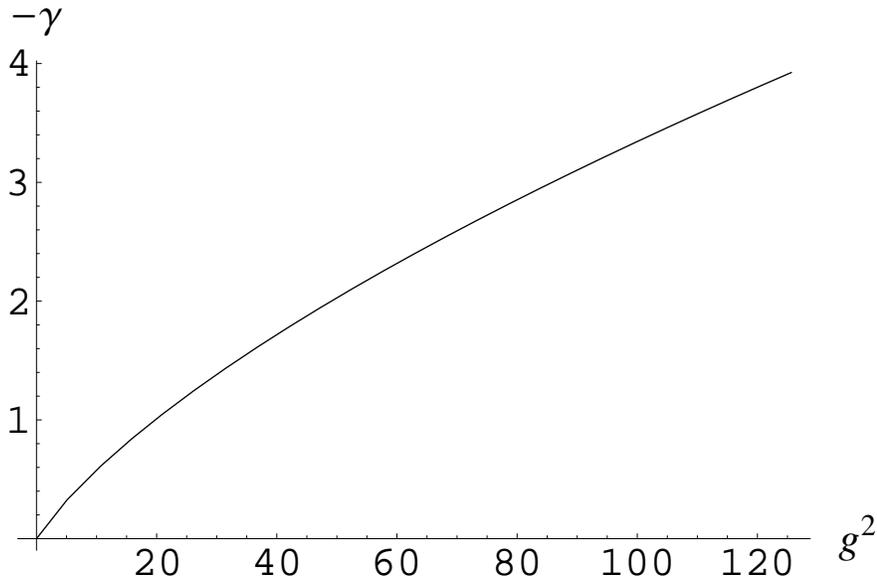}
\caption{\small Anomalous dimension as a function of $g^2$}
\label{F12T2Lat}
\end{figure}

\section{Discussion}
In this letter we have generalized the calculations of~\cite{BrKr01}
to a supersymmetric setting. We have chosen to study a Wess--Zumino
like model in which the vertex is not renormalized and this allowed
us to calculate  the order zero of the renormalization group
functions in a large $N$ expansion in a closed form.

Writing the Dyson--Schwinger equations for propagators   and
proposing an appropriate ansatz for the corresponding self-energies
we have found a non-linear differential equation for the anomalous
dimension (eq.~(\ref{nonlinear})) from which one can derive a
perturbative expansion for $\gamma$. Coefficients in such expansion
can be very easily evaluated up to very large orders and a
Pad\'e--Borel resummation allowed us to compute $\gamma(g)$ in a large
range of coupling constants with great accuracy.

In our calculations, the power of the Hopf algebra approach has not
be used explicitly. It is tempting to speculate, however, that the
special features of perturbation theory in supersymmetric models
should reflect in some particular structure within the Hopf algebra
approach. We intend to discuss this issue in a future work.

\vspace{1 cm}

\noindent\underline{Acknowledgments}

\noindent We would like to thank the Sociedad Cientifica Argentina
for hospitality.
 This work is partially supported by CONICET (PIP6160), ANPCyT (PICT 20204),
 CONICET/CNRS/PICS-3172, UNLP, UBA and CICBA  grants. M.B acknowledges CNRS support
 through his ``mise \`a disposition".

\end{document}